\documentclass[a4paper]{article}

\usepackage[margin=30mm]{geometry}

\author{Koki Hamada \and Dai Ikarashi \and Ryo Kikuchi \and Koji Chida}
\date{
NTT Social Informatics Laboratories\\[2ex]
December 17, 2021}

\usepackage{hamada}

\usepackage[ruled,linesnumbered,noend]{algorithm2e}

\usepackage{cleveref}

\usepackage{fancybox}

\newcommand{\com}[1]{}
\newcommand{\seq}[1]{(#1)}

\newcommand{\forj}{for $j \in [1,m]$}
\newcommand{\fori}{for $i \in [1,n]$}

\newcommand{\enc}[1]{[\![#1]\!]}

\newcommand{\cMINVAL}{MAX\_VALUE}
\newcommand{\cNULL}{NULL}

\renewcommand{\cMINVAL}{{\sf MIN\_VALUE}}
\renewcommand{\cNULL}{{\sf NULL}}

\usepackage{amsfonts}
\newcommand{\setcapital}[1]{\mathbb{#1}}

\newcommand{\setZ}{\setcapital{Z}}

\newcommand{\bOR}{\binaryoperatorname{{\textsf{OR}}}}

\newcommand{\encP}[1]{\langle\!\langle #1 \rangle\!\rangle}

\newcommand{\ftrainU}{\funcdef{DTTrainU}}

\newcommand{\fsplitU}{\funcdef{SplitU}}

\newcommand{\fgini}{\funcdef{ModifiedGini}}

\newcommand{\fsortp}{\funcdef{SortPerm}}
\newcommand{\fsort}{\funcdef{Sort}}

\newcommand{\fshuffle}{\funcdef{Shuffle}}

\newcommand{\fapply}{\funcdef{Apply}}

\newcommand{\fapplyinv}{\funcdef{ApplyInv}}

\newcommand{\fmax}{\funcdef{Max}}

\newcommand{\fvectmax}{\funcdef{VectMax}}

\newcommand{\fEQ}{\funcdef{EQ}}
\newcommand{\fLT}{\funcdef{LT}}
\newcommand{\fAdd}{\funcdef{Add}}
\newcommand{\fMul}{\funcdef{Mul}}
\newcommand{\fifelse}{\funcdef{IfElse}}

\newcommand{\fpsum}{\funcdef{PrefixSum}}
\newcommand{\fpsumr}{\funcdef{PrefixSumR}}
\newcommand{\fpsuminv}{\funcdefi{PrefixSum}}
\newcommand{\fpsumrinv}{\funcdefi{PrefixSumR}}

\newcommand{\fgbsum}{\funcdef{GroupSum}}

\newcommand{\fgbpsum}{\funcdef{GroupPrefixSum}}

\newcommand{\fgbmax}{\funcdef{GroupMax}}

\newcommand{\faws}{\funcdef{AttributewiseSplit}}

\newcommand{\fFirstOne}{\funcdef{GroupFirstOne}}

\newcommand{\feq}[2]{(#1 \stackrel{?}{=} #2)}
\newcommand{\flt}[2]{(#1 \stackrel{?}{<} #2)}

\newcommand{\fhead}{\funcdef{Head}}
\newcommand{\ftail}{\funcdef{Tail}}

\newcommand{\fLayerShrink}{\funcdef{Shrink}}
\newcommand{\fGroupSame}{\funcdef{GroupSame}}
\newcommand{\feval}{\funcdef{Eval}}

\newcommand{\fabb}{\mathcal{F}_{\rm ABB}}
\newcommand{\fenc}{\funcdef{Enc}}
\newcommand{\fdec}{\funcdef{Dec}}

\newcommand{\xa}{a}

\newcommand{\xav}{\vec{\xa}}
\newcommand{\xavE}{\enc{\xav}}

\newcommand{\xavi}[1]{\xav_{#1}}
\newcommand{\xaviE}[1]{\enc{\xavi{#1}}}

\newcommand{\xavk}[1]{\xav[#1]}
\newcommand{\xavkE}[1]{\enc{\xavk{#1}}}

\newcommand{\xb}{b}

\newcommand{\xbv}{\vec{\xb}}
\newcommand{\xbvE}{\enc{\xbv}}

\newcommand{\xbvj}[1]{\xbv^{#1}}
\newcommand{\xbvjE}[1]{\enc{\xbvj{#1}}}

\newcommand{\xbvjk}[2]{\xbvj{#1}[#2]}

\newcommand{\xc}{c}
\newcommand{\xcE}{\enc{\xc}}

\newcommand{\xcv}{\vec{\xc}}
\newcommand{\xcvE}{\enc{\xcv}}

\newcommand{\xcvk}[1]{\xcv[#1]}
\newcommand{\xcvkE}[1]{\enc{\xcvk{#1}}}

\newcommand{\xd}{d}

\newcommand{\xdv}{\vec{\xd}}
\newcommand{\xdvE}{\enc{\xdv}}

\newcommand{\xdvi}[1]{\xdv_{#1}}
\newcommand{\xdviE}[1]{\enc{\xdvi{#1}}}

\newcommand{\xe}{e}

\newcommand{\xei}[1]{\xe_{#1}}

\newcommand{\xeik}[2]{\xei{#1}[#2]}
\newcommand{\xeikE}[2]{\enc{\xeik{#1}{#2}}}

\newcommand{\xev}{\vec{\xe}}

\newcommand{\xevi}[1]{\xev_{#1}}
\newcommand{\xeviE}[1]{\enc{\xevi{#1}}}

\newcommand{\xevik}[2]{\xevi{#1}[#2]}
\newcommand{\xevikE}[2]{\enc{\xevik{#1}{#2}}}

\newcommand{\xf}{f}

\newcommand{\xfi}[1]{\xf_{#1}}
\newcommand{\xfiE}[1]{\enc{\xfi{#1}}}

\newcommand{\xfv}{\vec{\xf}}
\newcommand{\xfvE}{\enc{\xfv}}

\newcommand{\xfvk}[1]{\xfv[#1]}
\newcommand{\xfvkE}[1]{\enc{\xfvk{#1}}}

\newcommand{\xg}{g}

\newcommand{\xgj}[1]{\xg^{#1}}
\newcommand{\xgjE}[1]{\enc{\xgj{#1}}}

\newcommand{\xgv}{\vec{\xg}}
\newcommand{\xgvE}{\enc{\xgv}}

\newcommand{\xgvj}[1]{\xgv^{#1}}
\newcommand{\xgvjE}[1]{\enc{\xgvj{#1}}}

\newcommand{\xgvjk}[2]{\xgvj{#1}[#2]}
\newcommand{\xgvjkE}[2]{\enc{\xgvjk{#1}{#2}}}

\newcommand{\xgvk}[1]{\xgv[#1]}
\newcommand{\xgvkE}[1]{\enc{\xgvk{#1}}}

\newcommand{\xh}{h}

\newcommand{\xk}{k}

\newcommand{\xl}{l}
\newcommand{\xlE}{\enc{\xl}}

\newcommand{\xlj}[1]{\xl^{#1}}
\newcommand{\xljE}[1]{\enc{\xlj{#1}}}

\newcommand{\xm}{m}

\newcommand{\xn}{n}

\newcommand{\xp}{p}

\newcommand{\xpk}[1]{\xp[#1]}

\newcommand{\xpv}{\vec{\xp}}
\newcommand{\xpvE}{\enc{\xpv}}

\newcommand{\xpvi}[1]{\xpv_{#1}}
\newcommand{\xpviE}[1]{\enc{\xpvi{#1}}}

\newcommand{\xpvk}[1]{\xpv[#1]}
\newcommand{\xpvkE}[1]{\enc{\xpvk{#1}}}

\newcommand{\xq}{q}

\newcommand{\xqk}[1]{\xq[#1]}

\newcommand{\xqv}{\vec{\xq}}
\newcommand{\xqvE}{\enc{\xqv}}

\newcommand{\xqvk}[1]{\xqv[#1]}
\newcommand{\xqvkE}[1]{\enc{\xqvk{#1}}}

\newcommand{\xs}{s}

\newcommand{\xsv}{\vec{\xs}}
\newcommand{\xsvE}{\enc{\xsv}}

\newcommand{\xsvi}[1]{\xsv_{#1}}
\newcommand{\xsviE}[1]{\enc{\xsvi{#1}}}

\newcommand{\xsvik}[2]{\xsvi{#1}[#2]}
\newcommand{\xsvikE}[2]{\enc{\xsvik{#1}{#2}}}

\newcommand{\xsvk}[1]{\xsv[#1]}
\newcommand{\xsvkE}[1]{\enc{\xsvk{#1}}}

\newcommand{\xt}{t}

\newcommand{\xti}[1]{\xt_{#1}}
\newcommand{\xtiE}[1]{\enc{\xti{#1}}}

\newcommand{\xtv}{\vec{\xt}}
\newcommand{\xtvE}{\enc{\xtv}}

\newcommand{\xtvi}[1]{\xtv_{#1}}
\newcommand{\xtviE}[1]{\enc{\xtvi{#1}}}

\newcommand{\xtvik}[2]{\xtvi{#1}[#2]}
\newcommand{\xtvikE}[2]{\enc{\xtvik{#1}{#2}}}

\newcommand{\xtvk}[1]{\xtv[#1]}
\newcommand{\xtvkE}[1]{\enc{\xtvk{#1}}}

\newcommand{\xu}{u}

\newcommand{\xuv}{\vec{\xu}}
\newcommand{\xuvE}{\enc{\xuv}}

\newcommand{\xuvi}[1]{\xuv_{#1}}
\newcommand{\xuviE}[1]{\enc{\xuvi{#1}}}

\newcommand{\xuvik}[2]{\xuvi{#1}[#2]}

\newcommand{\xuvk}[1]{\xuv[#1]}

\newcommand{\xv}{v}

\newcommand{\xvv}{\vec{\xv}}

\newcommand{\xvvi}[1]{\xvv_{#1}}
\newcommand{\xvviE}[1]{\enc{\xvvi{#1}}}

\newcommand{\xw}{w}

\newcommand{\xwv}{\vec{\xw}}
\newcommand{\xwvE}{\enc{\xwv}}

\newcommand{\xwvi}[1]{\xwv_{#1}}
\newcommand{\xwviE}[1]{\enc{\xwvi{#1}}}

\newcommand{\xwvik}[2]{\xwvi{#1}[#2]}

\newcommand{\xwvk}[1]{\xwv[#1]}

\newcommand{\xx}{x}

\newcommand{\xxv}{\vec{\xx}}
\newcommand{\xxvE}{\enc{\xxv}}

\newcommand{\xxvi}[1]{\xxv_{#1}}
\newcommand{\xxviE}[1]{\enc{\xxvi{#1}}}

\newcommand{\xxvij}[2]{\xxvi{#1}^{#2}}
\newcommand{\xxvijE}[2]{\enc{\xxvij{#1}{#2}}}

\newcommand{\xxvik}[2]{\xxvi{#1}[#2]}
\newcommand{\xxvikE}[2]{\enc{\xxvik{#1}{#2}}}

\newcommand{\xxvj}[1]{\xxv^{#1}}
\newcommand{\xxvjE}[1]{\enc{\xxvj{#1}}}

\newcommand{\xxvjk}[2]{\xxvj{#1}[#2]}
\newcommand{\xxvjkE}[2]{\enc{\xxvjk{#1}{#2}}}

\newcommand{\xxvk}[1]{\xxv[#1]}
\newcommand{\xxvkE}[1]{\enc{\xxvk{#1}}}

\newcommand{\xy}{y}

\newcommand{\xyv}{\vec{\xy}}
\newcommand{\xyvE}{\enc{\xyv}}

\newcommand{\xyvi}[1]{\xyv_{#1}}
\newcommand{\xyviE}[1]{\enc{\xyvi{#1}}}

\newcommand{\xyvik}[2]{\xyvi{#1}[#2]}

\newcommand{\xyvj}[1]{\xyv^{#1}}
\newcommand{\xyvjE}[1]{\enc{\xyvj{#1}}}

\newcommand{\xyvk}[1]{\xyv[#1]}
\newcommand{\xyvkE}[1]{\enc{\xyvk{#1}}}

\newcommand{\xz}{z}
\newcommand{\xzE}{\enc{\xz}}

\newcommand{\xzi}[1]{\xz_{#1}}
\newcommand{\xziE}[1]{\enc{\xzi{#1}}}

\newcommand{\xzv}{\vec{\xz}}
\newcommand{\xzvE}{\enc{\xzv}}

\newcommand{\xzvi}[1]{\xzv_{#1}}
\newcommand{\xzviE}[1]{\enc{\xzvi{#1}}}

\newcommand{\xzvk}[1]{\xzv[#1]}
\newcommand{\xzvkE}[1]{\enc{\xzvk{#1}}}

\newcommand{\xA}{A}
\newcommand{\xAE}{\enc{\xA}}

\newcommand{\xAj}[1]{\xA^{#1}}
\newcommand{\xAjE}[1]{\enc{\xAj{#1}}}

\newcommand{\xAk}[1]{\xA[#1]}
\newcommand{\xAkE}[1]{\enc{\xAk{#1}}}

\newcommand{\xL}{L}

\newcommand{\xLj}[1]{\xL^{#1}}
\newcommand{\xLjE}[1]{\enc{\xLj{#1}}}

\newcommand{\xN}{N}
\newcommand{\xNE}{\enc{\xN}}

\newcommand{\xNj}[1]{\xN^{#1}}
\newcommand{\xNjE}[1]{\enc{\xNj{#1}}}

\newcommand{\xNjk}[2]{\xNj{#1}[#2]}
\newcommand{\xNjkE}[2]{\enc{\xNjk{#1}{#2}}}

\newcommand{\xP}{P}

\newcommand{\xPP}{\encP{\xP}}

\newcommand{\xS}{S}

\newcommand{\xT}{T}
\newcommand{\xTE}{\enc{\xT}}

\newcommand{\xTj}[1]{\xT^{#1}}
\newcommand{\xTjE}[1]{\enc{\xTj{#1}}}

\newcommand{\xTk}[1]{\xT[#1]}
\newcommand{\xTkE}[1]{\enc{\xTk{#1}}}

\newcommand{\xX}{X}

\newcommand{\xY}{Y}

\crefname{figure}{Fig.}{Figs.}
\crefname{table}{Table}{Tables}
\crefname{section}{Section}{Sections}

\crefname{algorithm}{Algorithm}{Algorithms}
\crefname{line}{Line}{Lines}

\crefname{ALC@unique}{Step}{Steps}
\crefname{equation}{Equation}{Equations}

\renewcommand{\xl}{{\sf Label}}
\renewcommand{\xN}{{\sf NID}}
\renewcommand{\xT}{{\sf Threshold}}
\renewcommand{\xA}{{\sf AID}}
\renewcommand{\xL}{{\sf Layer}}

\newcommand{\algorithmcommon}{
 \DontPrintSemicolon
 \SetKwInput{KwNotation}{Notation}
 \SetKwInput{KwNot}{Notation}
 \SetKwInput{KwComp}{Cost}
 \SetKwFor{ForEachDIP}{for each}{do in parallel}{end for}
}

\newenvironment{hoge}{}{}

\newcommand{\hamadaemph}[1]{\textbf{\underline{#1}}}
\newcommand{\IGV}{\hamadaemph{internally grouped vector}}
\newcommand{\GFV}{\hamadaemph{group flag vector}}
\newcommand{\GTS}{\hamadaemph{global test selection}}
\newcommand{\LTS}{\hamadaemph{attribute-wise test selection}}
\newcommand{\BR}{\hamadaemph{test result}}
\newcommand{\TEST}{\hamadaemph{test}}
\newcommand{\SCORE}{\hamadaemph{score}}
\newcommand{\PDS}{\hamadaemph{private grouping data structure}}
\newcommand{\LL}{\hamadaemph{leaf label}}
\newcommand{\MGI}{\hamadaemph{modified Gini index}}

\newcommand{\testa}[2]{\fbox{$\xX_{#1} < #2$}}
\newcommand{\testb}[2]{\fbox{$2\xX_{#1} < #2$}}

\renewcommand{\feval}{\funcdef{ApplyTests}}

\newcommand{\fLeafNode}{\funcdef{TrainLeafNodes}}
\newcommand{\fInNode}{\funcdef{TrainInternalNodes}}

\renewcommand{\fsplitU}{\funcdef{GlobalTestSelection}}
\renewcommand{\faws}{\funcdef{AttributewiseTestSelection}}
\renewcommand{\ftrainU}{\funcdef{DecisionTreeTraining}}
\renewcommand{\fLayerShrink}{\funcdef{FormatLayer}}
\renewcommand{\fapplyinv}{\funcdef{Unapply}}
\renewcommand{\encP}[1]{\enc{#1}}

\renewcommand{\hamadaemph}[1]{#1}

\renewcommand{\testa}[2]{$\xX_{#1} < #2$}
\renewcommand{\testb}[2]{$2\xX_{#1} < #2$}
\begin{document}
\title{\huge Efficient decision tree training with new data structure for secure multi-party computation}

\maketitle

\begin{abstract}
{
We propose a secure multi-party computation (MPC) protocol
that constructs a secret-shared decision tree for a given secret-shared dataset.
The previous MPC-based decision tree training protocol (\mbox{Abspoel} et al. 2021)
requires $O(2^hmn\log n)$ comparisons,
being exponential in the tree height $h$
and with $n$ and $m$
being the number of rows and that of attributes in the dataset, respectively.
The
cause of the exponential number of comparisons
in $h$ is that
the decision tree training algorithm is based on the divide-and-conquer paradigm,
where dummy rows are added after each split in order to hide the number of rows
in the dataset.
We
resolve this issue via secure data structure that
enables us to compute an aggregate value for
every group while hiding the grouping information.
By using this data structure,
we can train a decision tree without adding
dummy rows while hiding the size of the intermediate data.
We specifically
describes a decision tree training protocol
that requires only $O(hmn\log n)$ comparisons when the input attributes are continuous and the output attribute is binary.
Note that the order is now
\emph{linear} in the tree height~$h$.
To demonstrate the practicality of our protocol,
we implement it
in an MPC framework based on a three-party secret sharing scheme.
Our implementation results show that our protocol trains a decision tree with a height of 5 in 33 seconds for a dataset of 100,000 rows and 10 attributes.
}
\end{abstract}

\section{Introduction}

\begin{hoge}
Secure multi-party computation (MPC)~\cite{Yao86}
allows parties to jointly compute any function while keeping inputs private.
Its large computational overhead has long been a barrier to practical use.
In recent years, even efficient MPC protocols for machine learning methods
such as neural network training~\cite{WGC19,RWT+18,MR18} have been proposed.

Decision tree is one of the classical machine learning methods.
It is still widely used due to its computational simplicity and ease of interpretation.
It is also an important component of other machine learning methods,
such as gradient boosting decision tree~\cite{Fri01} and random forest~\cite{Bre01},
which have been successful in recent years.

Since the work of Lindell and Pinkas~\cite{LP00}
in the early days of privacy-preserving data mining,
there has been a lot of research on MPC protocols for decision tree training.
In order to be used as a component of MPC protocols for other machine learning methods,
it is desirable to keep all the information, from the input to the trained decision tree, private.
However, only a few protocols~\cite{HSCA14,AEV21,ACC+21} with such a property have been proposed.
This is mainly due to two kinds of computational difficulties in MPC.

The first difficulty is the computation on real numbers.
Decision tree training requires computation of evaluation functions.
Although there are many types of evaluation functions, all commonly used ones involve division or logarithm.
Therefore, naive MPC protocols for decision tree training involve computation on real numbers, which increases the computational cost.
On the contrary, de Hoogh et al.~\cite{HSCA14} cleverly avoided computation
on real numbers by replacing fractional number comparisons with integer comparisons,
and proposed an efficient protocol for the case where inputs are categorical values.
Abspoel et al.~\cite{AEV21} presented an efficient protocol that can be applied to the case where the input contains numerical values.
The number of candidates to which the evaluation functions are applied is $\Theta(c)$
when the input consists only of categorical values,
whereas it increases to $\Theta(n^2)$ when the input contains numerical values,
where $c$ is the number of possible values of the categorical value and $n$ is the number of samples in the input.
They used a sorting protocol to reduce the number of candidates to $O(n)$,
and also extended the technique of de Hoogh et al. to the numerical case to avoid computation on real numbers.
Adams et al.~\cite{ACC+21} dealt with the case where the input contains numeric values
by a different approach: discretizing the numeric attributes of the input.
Although the trained tree is slightly different from the one without discretization,
this approach avoids the use of sorting, which is relatively computationally expensive,
and allows us to use the efficient protocol of de Hoogh et al.~\cite{HSCA14}.

The second difficulty is the protection of the intermediate data size.
In decision tree training, the data is split recursively from the root node to the leaf nodes in a top-down fashion.
As the tree height increases, the number of nodes increases exponentially.
On the other hand, the size of the intermediate data processed by each node also decreases exponentially on average,
hence the overall computational cost is linear in the tree height.
When this is implemented in MPC, the intermediate data size after splitting has to be hidden,
so the existing protocols~\cite{HSCA14,AEV21,ACC+21} used a data, which contains some dummy entries, of the same size as the original one at each node.
Therefore, we could not benefit from the size reduction by data splitting, and as a result, the overall computational cost was exponential in the tree height.
\end{hoge}

\subsection{Our contribution}

\begin{table}
 \centering
 \caption{Comparison of computational cost of MPC protocols for decision tree training with numerical input attributes.}
 \label{tbl:comparing-comm}
 \begin{tabular}{ll}
  Method & Number of operations \\
  \hline 
  Trivial~\cite{AEV21} & $O(2^h m n^2)$ \\ 
  Abspoel et al.~\cite{AEV21} & $O( (2^h + \log n) m n \log n )$  \\ 
  Abspoel et al.~\cite{AEV21}& $O( 2^h m n \log n )$  \\ 
   with efficient sort~\cite{HKI+12}&   \\ 
  Ours & $O(h m n \log n)$ \\
 \end{tabular}
\end{table}

We propose an MPC protocol for decision tree training with linear computational cost on tree height, which is the first protocol to solve the second problem above.
It trains a binary decision tree under the assumption that all input attributes are numeric and the output attribute is binary.
As in the protocol by Abspoel et al.~\cite{AEV21}, it does not reveal any information other than the size of the input and the upper bound $h$ on the tree height.

The computational cost of our protocol is $O(h m n \log n)$, assuming that the comparison and multiplication protocols are unit operations,
where $m$ is the number of input attributes in the dataset, and $n$ is the number of samples in the dataset.
This is an exponential improvement with respect to $h$ over the computational cost $O(2^h m n \log n)$ of the protocol by Abspoel et al.
(Actually, Abspoel et al.~\cite{AEV21} claimed only a computational cost $O((2^h+\log n) m n \log n)$,
but their protocol can easily be implemented to run in $O(2^h m n \log n)$ by replacing the sorting protocol to efficient one such as~\cite{HKI+12}.)
A comparison of the computational costs is shown in \cref{tbl:comparing-comm}.

Our approach of exponential improvement in computational cost with respect to the tree height is general.
For completeness, our protocol is instantiated with all input attributes being numeric,
the output attribute being binary, and the evaluation function being the Gini index; however, it is easy to extend.
In fact, the main protocol (\cref{alg:train}),
which plays a central role in the exponential improvement of the computational cost,
describes a process common to the major decision tree training algorithms CART~\cite{BFOS84}, ID3~\cite{Qui86}, and C4.5~\cite{Qui14}.

Our protocol is built on top of a set of basic protocols,
such as multiplication and comparison, provided by many recent MPC frameworks, so it can be used on top of various implementations.
More specifically, we build our protocol on top of an MPC model called arithmetic black box (ABB),
which consists of a set of basic operations described in \cref{sec:abb}.

As a byproduct of our decision tree training protocol,
we also propose a secure data structure that can compute aggregate values
for each group while keeping the grouping information private.
This data structure can be used to compute aggregate values such as sums and maximums within each group
while keeping the grouping information private, even in cases other than decision tree training.

To see the practicality of our decision tree training protocol, we implemented it on an MPC framework based on a 2-out-of-3 secret sharing scheme.
Our protocol trained a decision tree with a height of 5 for 100,000 inputs of 10 input variables in 33 seconds.

\subsection{Overview of our techniques}
In general, MPC protocols are incompatible with divide-and-conquer algorithms.
In divide-and-conquer algorithms, the problem is divided into smaller subproblems and solved recursively, but MPC protocols need to hide the size of the divided problem as well.
A common way to hide the size of the problem is to add a dummy.
We hide the actual size of the data by adding dummies (or leaving samples that should be removed) to the split data to make it appear to be the same size as the original.
A disadvantage of this method is that it is computationally expensive; since it loses the property that the data size becomes smaller after splitting.
For this reason, the previous study~\cite{AEV21} required an exponential cost for the height of the tree.

We use the property that the total number of samples is invariant at each height in training decision trees.
We keep the data of nodes of the same height together, and train them all at once without adding any dummies.
This allows our protocol to process only $\Theta(h m n)$ samples in total,
while the previous study~\cite{AEV21} processes $\Theta(2^h m n)$ samples including dummies.

To implement this idea, we first define a data structure that looks like a private vector of length $n$, but is internally grouped.
Specifically, we place the $n$ grouped elements on a private vector of length $n$ so that elements of the same group appear next to each other, and then create a private vector of length n with a flag corresponding to the first element of each group.
This allows us to detect the boundaries of groups by referring to the flags internally, although we cannot distinguish the groupings outwardly.

In decision tree training, each group needs to be split when moving to the next height.
We accomplish this within our data structure by stably sorting the elements using the binary branching result, which is computed for each element, as a key.
Stability of the sort ensures that elements that are in the same group and have the same branching result will be placed sequentially after the sort.
Since this split requires only one-bit-key sorting, it can be very efficient depending on the underlying MPC implementation.

We build the group-wise sum, maximum, and prefix sum computations on our data structure.
We then use them to build a decision tree training algorithm similar to~\cite{AEV21} on our data structure.

\section{Preliminaries}
In this section, we introduce a typical decision tree training algorithm in the clear and secure mulit-party computation.

Before that, we introduce some notation.
Throughout this paper, the index of a vector starts at $1$.
We refer to the $i$-th element of a vector $\xvv$ by $\xvv[i]$.
That is, if $\xvv$ is a vector of length $n$, then $\xvv = (\xvv[1], \xvv[2], \dots, \xvv[n])$.
In logical operations, $0$ represents false and $1$ represents true.

\subsection{Decision tree training}
\newcommand{\hX}[1]{\funcdef{H_{#1}}}
\newcommand{\pTrain}{\funcdef{Train}}
\newcommand{\pGini}[1]{\funcdef{Gini_{\mathit{#1}}}}
\newcommand{\pGain}[1]{\funcdef{Gain_{\mathit{#1}}}}
\newcommand{\pG}[1]{\funcdef{G_{\mathit{#1}}}}
\newcommand{\pGp}[1]{\funcdef{G'_{\mathit{#1}}}}
\newcommand{\pTesta}{\xX_j <   t}
\newcommand{\pTestb}{\xX_j \ge t}
\newcommand{\Dxx}{\mathcal{D}}
\newcommand{\Dxa}{\Dxx_{\xY=0}}
\newcommand{\Dxb}{\Dxx_{\xY=1}}
\newcommand{\Dxc}{\Dxx_{\xY=c}}
\newcommand{\Dax}{\Dxx_{\pTesta}}
\newcommand{\Dbx}{\Dxx_{\xX_j \ge t}}
\newcommand{\Daa}{\Dxx_{\pTesta \wedge \xY = 0}}
\newcommand{\Dba}{\Dxx_{\xX_j \ge t \wedge \xY = 0}}
\newcommand{\Dab}{\Dxx_{\pTesta \wedge \xY = 1}}
\newcommand{\Dbb}{\Dxx_{\xX_j \ge t \wedge \xY = 1}}
\newcommand{\Dac}{\Dxx_{\pTesta \wedge \xY = c}}
\newcommand{\Dbc}{\Dxx_{\xX_j \ge t \wedge \xY = c}}
\newcommand{\Tx}{\mathcal{T}}
\newcommand{\Ta}{\Tx_{\pTesta}}
\newcommand{\Tb}{\Tx_{\pTestb}}

 Decision tree training is a method in machine learning.
 The goal is to obtain a model called a {\em decision tree}
 that predicts a value of an {\em output attribute},
 given values of {\em input attributes}.
There are several famous algorithms for decision tree training, such as CART~\cite{BFOS84}, ID3~\cite{Qui86}, and C4.5~\cite{Qui14}.
The general framework of these algorithms is the same, and in fact they are all greedy algorithms based on the divide-and-conquer paradigm.
In this section, we present a typical algorithm, for which we plan to construct a secure version, for training a two-class classification binary tree,
where all input attributes are numerical.

\subsubsection{Typical decision tree training algorithm}
\label{sec:typical-dt-train}
 Let us start with defining notation.
 Consider a dataset $D$ with
 $m$ input attributes $X_1,\dots,X_m$ and
 an output attribute $Y$.
Suppose there are $\xn$ {\em samples},
each sample being a pair of an {\em input tuple} $x$ and a {\em class label} $y$.
 Here, $x$ is an $m$-tuple,
and $y$ is a value of the output attribute $Y$.
The $j$-th element of $x$ represents a value of the input attribute $X_j$.
 A decision tree consists of a binary tree and some additional information.
 Each internal node (non-leaf node) has a condition called a {\em test} of the form $\pTesta$.
 It asks if the $j$-th element in a given input tuple is less than a {\em threshold} $t$ or not.
 Each edge is assigned a possible outcome of its source node's test, that is, true or false.
 An edge whose assigned outcome is true (false) is called a {\em true edge} ({\em false edge}, respectively).
 A child node whose incoming edge is a true edge (false edge) is called a {\em true-child node} ({\em false-child node}, respectively).
 Each leaf node is assigned a class label called {\em \LL{}}.
 This information is used to predict a class label for a given input tuple as follows.
 Starting from the root node, we repeat evaluating the test of the internal node we reach and
 tracing an outgoing edge that is assigned the same value as the test outcome.
 When we reach a leaf node, we output its \LL{} as the predicted class label.

\begin{algorithm}[htb]
 \caption{A typical decision tree training algorithm in the clear.}
 \label{alg:dt-train-plain}
 \DontPrintSemicolon
 \SetKwInput{KwNotation}{Notation}
 \KwNotation{$\Tx := \pTrain{\Dxx}$}
 \KwIn{A training dataset $\Dxx$.}
 \KwOut{A decision tree $\Tx$.}
 \eIf{the stopping criterion is met}{
    Let $r$ be a leaf node whose \LL{} is the most common class label in $\Dxx$.
    Outputs a tree whose root node is $r$.
 }{
    Find the best test $\pTesta$ according to the variable selection measure.
    \label{step:dt-train-plain-split}
    \;
    Recursively computes two subtrees $\Ta := \pTrain{\Dax}$ and $\Tb := \pTrain{\Dbx}$. \;
    Let $v$ be an internal node $v$ whose test is $\pTesta$.
    Output a tree such that its root node is $v$,
    $v$'s true-child node is $\Ta$'s root node, and
    $v$'s false-child node is $\Tb$'s root node. \;
 }
\end{algorithm}

A typical decision tree training algorithm is shown in \cref{alg:dt-train-plain}.
It trains a tree recursively from the root node to a leaf node in a top-down fashion.
At each node, it checks if the stopping criterion is satisfied using the given training dataset $\Dxx$ to determine the node type.
If the stopping criterion is satisfied, the current node is set to be a leaf node.
Then, it sets the most frequent class label in the dataset to the \LL{} of the current node, and outputs a tree whose root is this node.
If the stopping criterion is not satisfied, the current node is set as an internal node.
In this case, we select a test of the form $\pTesta$ that gives the best data splitting
with respect to a predetermined criterion, and split the training dataset $\Dxx$ into $\Dax$ and $\Dbx$ according to this test,
where
$\Dax := \set{ (x,y) \in \Dxx \mid x(X_j) < t }$
and
$\Dbx := \set{ (x,y) \in \Dxx \mid x(X_j) \ge t }$.
It then recursively trains decision trees $\Ta$ and $\Tb$ with $\Dax$ and $\Dbx$ as the training data, respectively,
and sets the roots of these trees as the child nodes of the current node, and outputs a tree whose root is the current node.

We use the commonly used stopping criterion: (1) the height of the node is $h$, or (2) the dataset cannot be split further (i.e., (i) all class labels are the same, or (ii) all input tuples are the same),
where $h$ is an upper bound of the tree height, which is typically given as a hyperparameter.

\subsubsection{Test selection measure}
The size and shape of the decision tree depends on which tests are selected at the internal nodes.
In general, it is desirable to make the tree as small as possible, but the problem of constructing a decision tree that minimizes the sum of the lengths of the paths from the root to each leaf is known to be NP-hard \cite{HR76}.
Therefore, we usually define a measure for goodness of local splitting and select a test that maximizes this measure.

Commonly used measures for goodness of split include the information gain used in ID3 \cite{Qui86} and the Gini index used in CART \cite{BFOS84}.
We use the Gini index, which is also used in previous studies such as \cite{HSCA14,AEV21} due to its ease of computation in MPC.

Two types of Gini indices are defined: one for a dataset and one for a dataset and a test.
The Gini index for a dataset $\Dxx$, which we denote by $\pGini{}{\Dxx}$, is defined as follows:
\[
 \pGini{}{\Dxx} := 1 - \sum_{c\in\set{0,1}} \frac{|\Dxc|^2}{|\Dxx|^2},
\]
where
$\Dxc := \set{ (x,y) \in \Dxx \mid y=c }$ is a subset of $\Dxx$ whose class label is $c$.
Intuitively, the smaller $\pGini{}{\Dxx}$ is, the purer $\Dxx$ becomes in terms of class labels.

The Gini index for a dataset $\Dxx$ and a test $\pTesta$,
which we denote by $\pG{\pTesta}{\Dxx}$,
is defined using $\pGini{}{}$ as follows:
\[
\pG{\pTesta}{\Dxx} := \frac{|\Dax|}{|D|} \pGini{}{\Dax} + \frac{|\Dbx|}{|\Dxx|} \pGini{}{\Dbx}.
\]
Intuitively, the smaller $\pG{\pTesta}{\Dxx}$ is, the purer each split dataset becomes (and hence the better the test is).
Therefore, to find the best test for splitting a dataset $\Dxx$,
we compute a test $T$ that minimizes $\pG{T}{\Dxx}$ \cite{HKP11}.

Abspoel et al.~\cite{AEV21} showed that minimization of $\pG{\pTesta}{\Dxx}$
is equivalent to maximization of $\pGp{\pTesta}{\Dxx}$ defined as
\begin{multline}
 \pGp{\pTesta}{\Dxx} := ( |\Dbx|( |\Daa|^2 + |\Dab|^2 ) \\  + |\Dax|( |\Dba|^2 + |\Dbb|^2) ) \\  / (|\Dax| |\Dbx|), \label{eq:modified-gini}
\end{multline}
where
$\Dac := \set{ (x,y) \in \Dxx \mid x(X_j) < t \wedge y=c }$ and
$\Dbc := \set{ (x,y) \in \Dxx \mid x(X_j) \ge t \wedge y=c }$.
We refer to it as {\em modified Gini index} and use it as a measure in our protocol.

\subsection{Secure multi-party computation}
We model secure multi-party computation (MPC) with an ideal functionality called arithmetic black box (ABB).
This ideal functionality allows a set of parties
$P_1,\dots,P_C$
to store values, operate on the stored values, and retrieve the stored values.
We build our protocol on top of an ABB.
This allows our protocol to run on any MPC implementation that realizes ABB, since concrete ABB implementation is separated from their construction.

\subsubsection{Arithmetic black box}\label{sec:abb}

\begin{figure}
 \centering
\fbox{
  \begin{minipage}{0.92\columnwidth}
 \begin{itemize}
  \item A command $\enc{z} \gets \fenc{x, P_i}$:
	Receive $x$ from a party $P_i$ and store it as $\enc{x}$.
  \item A command $z \gets \fdec{\enc{x}}$:
	Send $x$ to every party, who store it in the local variable $z$.
  \item A command $\enc{z} \gets \fAdd{\enc{x}, \enc{y}}$:
	Compute $z := x + y$ and store it as $\enc{z}$.
  \item A command $\enc{z} \gets \fMul{\enc{x}, \enc{y}}$:
	Compute $z := xy$ and store it as $\enc{z}$.
  \item A command $\enc{z} \gets \fEQ{\enc{x}, \enc{y}}$:
	If $x = y$ then set $z := 1$, otherwise set $z := 0$.
	Store it as $\enc{z}$.
  \item A command $\enc{z} \gets \fLT{\enc{x}, \enc{y}}$:
	If $x < y$ then set $z := 1$, otherwise set $z := 0$.
	Store it as $\enc{z}$.
 \end{itemize}
 We assume that the commands $\fAdd{}$, $\fMul{}$, $\fEQ{}$, and $\fLT{}$
 are also defined in the same way when one of the inputs is a public value.
  \end{minipage}
}
 \caption{The arithmetic black box functionality $\fabb$.}
 \label{fig:fabb}
\end{figure}

We assume a simple ABB named $\fabb$ over a ring $\setZ_M$ for some integer $M$ as shown in \cref{fig:fabb}.
We denote a value referred to by a name $x$ stored in $\fabb$ as $\enc{x}$. 
In a typical case, where $\fabb$ is realized by a secret sharing based MPC, $\enc{x}$ means that $x$ is secret shared.
We say a value is {\em private} if it is stored in $\fabb$.

We identify residue classes in $\setZ_M$ with their representatives in $[0,M)$.
We assume $M$ is sufficiently large such that vector indices can be stored in $\fabb$.
We also assume that the number of parties $C$ is constant.

For notational simplicity,
 $\enc{z} \gets \fAdd{\enc{x}, \enc{y}}$,
 $\enc{z} \gets \fMul{\enc{x}, \enc{y}}$,
 $\enc{z} \gets \fEQ{\enc{x}, \enc{y}}$, and
 $\enc{z} \gets \fLT{\enc{x}, \enc{y}}$
are also written as
 $\enc{z} \gets \enc{x} + \enc{y}$,
 $\enc{z} \gets \enc{x} \times \enc{y}$,
 $\enc{z} \gets \feq{ \enc{x} }{ \enc{y} }$, and
 $\enc{z} \gets \flt{ \enc{x} }{ \enc{y} }$, respectively.
Furthermore, we denote
$\enc{x_1} + \enc{x_2} + \dots + \enc{x_n}$
by $\sum_{i=1}^{n} \enc{x_i}$.

\subsubsection{Cost of MPC protocols}
We define the cost of an MPC protocol as the number of
invocations of ABB operations
other than linear combination of private values.
That is, we assume that the parties can compute
$\fAdd{\enc{x}, \enc{y}}$,
$\fAdd{c, \enc{y}}$,
$\fAdd{\enc{x}, c}$,
$\fMul{c, \enc{y}}$, and
$\fMul{\enc{x}, c}$ for free,
where $\enc{x}$ and $\enc{y}$ are private values and $c$ is a public value.
This cost models the communication complexity on a typical MPC
based on a linear secret sharing scheme,
in which the parties can locally compute linear combination of secret shared values.
We refer to ABB operations, except for linear combinations of private values,
as {\em non-free operations}.

\subsubsection{Known protocols}
We show known protocols that we will use as building blocks for our protocols.
The protocols shown here are limited to those that can be built on $\fabb$ for completeness.
Some MPC implementations may provide the same functionality more efficiently,
in which case we can use them instead of the protocols listed here to run our protocol more efficiently.

We start by defining some simple protocols.
\begin{itemize}
 \item $\enc{z} \gets \enc{x} \bOR \enc{y}$
       computes logical disjunction of bits $x$ and $y$ as $\enc{x} + \enc{y} - \enc{x} \times \enc{y}$,
       using $O(1)$ non-free operations in $O(1)$ rounds.
 \item $\enc{z} \gets \lnot\enc{x}$
       computes negation of a bit $x$ as $1 - \enc{x}$,
       using no non-free operations.
 \item $\enc{z} \gets \fifelse{\enc{c}, \enc{t}, \enc{f}}$
       receives a bit $c$ and two values $t$ and $f$, and
       computes $t$ if $c = 1$, $f$ otherwise, as $\enc{f} + \enc{c} \times (\enc{t} - \enc{f})$,
       using $O(1)$ non-free operations in $O(1)$ rounds.
 \item $\enc{z} \gets \fmax{\enc{x}, \enc{y}}$
       computes the maximum value of $x$ and $y$ as $\fifelse{ \flt{ \enc{x} }{ \enc{y} }, \enc{y}, \enc{x} }$,
       using $O(1)$ non-free operations in $O(1)$ rounds.
\end{itemize}

We require an extended $\fmax{}$ protocol, which we call $\fvectmax{}$.
We let $\xzE \gets \fvectmax{ \xxvE, \xyvE }$ denote the operation that
computes a private value $\xzE$ such that
$\xxvE$ and $\xyvE$ are private vectors of the same length $\xn$,
$i = \argmax_{j\in[1,\xn]} \xxvk{j}$, and $\xz = \xyvk{j}$.
We use the construction by Abspoel et al.~\cite{AEV21},
which uses $O(\xn)$ non-free operations in $O(\log n)$ rounds.

We require three permutation-related protocols $\fshuffle{}$, $\fapply{}$, and $\fapplyinv{}$.
Let $S_n$ be a symmetric group on $[1,n]$.
That is, $S_n$ is the set of all bijective functions from $[1,n]$ to $[1,n]$.
A {\em permutation} is an element of $S_n$.
{\em Applying} a permutation $\pi \in S_n$ to a vector $\xxv$ of length $n$ is the operation of rearranging $\xxv$
into a vector $\xzv$ satisfying $\xzv[\pi(i)] = \xxv[i]$ for $i\in[1,n]$.
We denote this operation as $\pi(\xxv)$.
Now we define the three protocols.
Let $\encP{\pi} \gets \fshuffle{n}$,
where $n$ is an integer,
denote the operation that computes $\encP{\pi}$,
such that $\pi$ is a uniformly randomly chosen element of $S_n$.
Let $\xzvE \gets \fapply{ \encP{\pi}, \xxvE }$,
where $\pi \in S_n$ is a permutation and $\xxv$ is a vector of length $n$,
denote the operation that computes $\xzvE$,
such that $\xzvE = \pi(\xxv)$.
Let $\xzvE \gets \fapplyinv{ \encP{\pi}, \xxvE }$,
where $\pi \in S_n$ is a permutation and $\xxv$ is a vector of length $n$,
denote the operation that computes $\xzvE$,
such that $\xzvE = \pi^{-1}(\xxv)$.
We use the construction by Falk and Ostrovsky \cite{FO21}.
In their construction, a private permutation is represented as
a set of private control bits for Waksman permutation network \cite{Wak68}.
All these protocols use $O(n \log n)$ non-free operations in $O(\log n)$ rounds.
Note that we do not use the $\fshuffle{}$ protocol directly in our protocols,
but it is required as a component of the construction of
$\fsortp{}$ protocol shown below.

We also require a protocol $\fsortp{}$ to compute a permutation that stably sorts given keys.
We let
\[
 \encP{\pi}\gets\fsortp{\xxviE{1}, \xxviE{2}, \dots, \xxviE{k}}
\]
denote the operation that computes $\encP{\pi}$
such that
$\xxvi{1}, \xxvi{2}, \dots, \xxvi{k}$ are vectors of length $n$ and
applying $\pi\in S_n$ to $((\xxvi{1}[1], \dots, \xxvi{k}[1]), \dots, (\xxvi{1}[n], \dots, \xxvi{k}[n]))$
lexicographically and stably sorts them.
We use the construction by Laud and Willemson \cite{LW14}.
The protocol use $O(n \log n)$ non-free operations in $O(\log n)$ rounds.
Note that we can construct the composition of private and public permutations
that is needed for the $\fsortp{}$ construction,
since our private permutation is a set of control bits for Waksman permutation network.

To simplify the description, we introduce a small subprotocol $\fsort{}$ for sorting private vectors.
We let
\[
 \xzviE{1}, \dots, \xzviE{\xm} \gets \fsort{ \xxviE{1}, \dots, \xxviE{\xk}; \xyviE{1}, \dots, \xyviE{\xm} }
\]
denote the following procedure:
\begin{enumerate}
 \item $\encP{\pi} \gets \fsortp{ \xxviE{1}, \allowbreak \dots, \allowbreak \xxviE{\xk} }$;
 \item $\xzviE{j} \gets \fapply{ \allowbreak \encP{\pi}, \allowbreak \xyviE{j} }$ for $j \in [1,\xm]$.
\end{enumerate}

We sometimes use similar notation when the same operation is applied to multiple inputs.
For example, 
\[
 \xziE{1}, \dots, \xziE{\xm} \gets \fifelse{\xcE; \xtiE{1}, \dots, \xtiE{\xm}; \xfiE{1},\dots,\xfiE{\xm}}
\]
means parallel execution of
$\xziE{j} \gets \fifelse{ \xcE, \xtiE{j}, \xfiE{j} }$
for $j \in [1,\xm]$ and 
\[
 \xziE{1}, \dots, \xziE{\xm} \gets \fvectmax{ \xxvE; \xyviE{1}, \dots, \xyviE{\xm} }
\]
means parallel execution of
$\xziE{j} \gets \fvectmax{ \xxvE, \xyviE{j} }$
for $j\in[1,\xm]$.

If vectors are given for a protocol defined for scalar values,
it means that the protocol is applied on an element-by-element basis.
That is,
$\xzvE \gets \xxvE \times \xyvE$
means parallel execution of
$\xzvkE{i} \gets \xxvkE{i} \times \xyvkE{i}$ for $i \in [1,\xn]$,
and
$\xzvE \gets \fifelse{ \xcvE, \xtvE, \xfvE }$
means parallel execution of
$\xzvkE{i} \gets \fifelse{ \xcvkE{i}, \xtvkE{i}, \xfvkE{i} }$ for $i \in [1,\xn]$.

If some of the inputs are scalar, the same scalar values are used for all executions.
For example,
$\xzvE \gets 2 \times \xyvE$
means parallel execution of
$\xzvkE{i} \gets 2 \times \xyvkE{i}$ for $i \in [1,\xn]$,
and
$\xzvE \gets \fifelse{ \xcE, \xtvE, 1 }$
means parallel execution of
$\xzvkE{i} \gets \fifelse{ \xcE, \xtvkE{i}, 1 }$ for $i \in [1,\xn]$.

\section{Our secure group-wise aggregation protocols}
\label{sec:group-ops}
In this section, we propose group-wise aggregation protocols
that compute aggregate values (sum, prefix sum, and maximum)
for each group without revealing the grouping information of the input grouped values.
These are executed on grouped values stored in our data structure.
These protocols and the data structure play a central role
in the construction of our decision tree training protocol proposed in \cref{sec:our-dt-training}.

\subsection{Our data structure for privately grouped values}
We propose a data structure that stores grouped values without revealing any information about the grouping.
We store $n$ values, divided into several groups, in a private vector $\xxvE$ of length $n$, called the {\em \IGV{}}.
Here, elements in the same group are stored as consecutive elements in the vector.
That is, for any $i$, $j$, and $k$ such that $1 \le i < j < k \le n$,
if $\xxv[i]$ and $\xxv[k]$ are in the same group, then $\xxv[i]$ and $\xxv[k]$ are also in the same group.
Along with such a vector, we maintain a private bit vector $\xgvE$ of length $n$, called the {\em \GFV{}},
which indicates the boundaries between groups.
Namely, we set $\xgvk{i} = 1$ if the $i$-th element in $\xxv$ is the first element in a group, otherwise $\xgvk{i} = 0$.
By definition, $\xgvk{1} = 1$ is always true.

We show an example.
Suppose that six values are stored in an \IGV{} $\xxv$ as $\xxv = (3,1,2,2,3,2)$
and the corresponding \GFV{} is $\xgv = (1,0,1,1,0,0)$.
Then, this means that the six values are divided into three groups, $(3,1)$, $(2)$, and $(2,3,2)$.

For the sake of simplicity, we introduce some notations.
Let $\fhead{\xgv, i}$ ($\ftail{\xgv, i}$) be the index of the first (last, respectively) element
of the group that contains the $i$-th element within the grouping represented by a \GFV{} $\xgv$.
Formally, they are defined as
$\fhead{\xgv, i} := \max \set{j \in [1,i] \mid \xgv[j] = 1}$ and
$\ftail{\xgv, i} := \min \set{j \in (i,n] \mid \xgv[j] = 1} \cup \set{n+1} - 1$, respectively,
where $n$ is the length of $\xgv$.
For example,
if a \GFV{} is defined as $\xgv = (1,0,1,1,0,0)$,
then $\fhead{\xgv, 2} = 1$, $\fhead{\xgv, 4} = 4$, $\ftail{\xgv, 4} = 6$, and $\ftail{\xgv, 3} = 3$.

\subsection{Our protocol for group-wise sum}

\begin{table}[tb]
\caption{Example of input/output for our group-wise aggregation protocols,
 where $\xgv$ is a \GFV{} and $\xxv$ is an \IGV{}.
}
 \centering
  \begin{tabular}{cc|ccc}
  \multicolumn{2}{c|}{Input} & \multicolumn{3}{c}{Output} \\
  \hline
  $\xgv$ & $\xxv$ & Sum & Prefix sum & Max\\
  \hline
  \hline
  1 & 3  & 4 & 3 & 3\\
  0 & 1  & 4 & 4 & 3\\
  \hline
  1 & 2  & 2 & 2 & 2\\
  \hline
  1 & 2  & 7 & 2 & 3\\
  0 & 3  & 7 & 5 & 3\\
  0 & 2  & 7 & 7 & 3\\
 \end{tabular}
 \label{fig:group-wise-io}
\end{table}

\begin{algorithm}[htb]
 \caption{Group-wise sum.}
 \label{alg:gb-sum}
 \algorithmcommon
 \renewcommand{\xP}{\pi}
 \KwNotation{$\xyvE \gets \fgbsum{\xgvE, \xxvE}$.}
 \KwIn{A private \GFV{} $\xgvE$ of length $\xn$ and
       a private \IGV{} $\xxvE$ of length $\xn$.}
 \KwOut{A private vector $\xyvE$ of length $\xn$.}
 \KwComp{$O(\xn \log \xn)$ non-free operations in $O(\log \xn)$ rounds.}

 $\xpvE \gets \fpsumr{\xxvE} \times \xgvE$.
   \label{step:gb-sum-psumr}
 \;
 $\xPP \gets \fsortp{\lnot\xgvE}$.
 \;
 $\xpviE{1} \gets \fapply{\xPP, \xpvE}$.
 \;
 $\xsviE{1} \gets \fpsumrinv{\xpviE{1}}$.
   \label{step:gb-sum-psumri}
 \;
 $\xdviE{1} \gets \fpsuminv{\xsviE{1}}$.
   \label{step:gb-sum-copy-psuminv}
 \;
 $\xdvE \gets \fapplyinv{\xPP, \xdviE{1}} \times \xgvE$.
   \label{step:gb-sum-copy-unapply}
 \;
 $\xyvE \gets \fpsum{\xdvE}$.
   \label{step:gb-sum-copy-psum}
 \;
\end{algorithm}

The {\em group-wise sum} protocol privately computes sums of each group in our data structure.
It receives a private \GFV{} $\xgvE$ of length $n$
and a private \IGV{} $\xxvE$ of length $n$,
and outputs a private vector $\xyvE$ of length $n$,
where $\xyv[i] = \sum_{j = \fhead{\xgv,i}}^{\ftail{\xgv,i}} \xxv[j]$ for $i \in [1,n]$.
Note that the same value is computed for elements in the same group.
An example is shown in \cref{fig:group-wise-io}.
Columns 1 and 2 are the inputs, and column 3 is the output.

Before presenting our protocol,
let us define some operations related to the computation of prefix sum.
Given a vector $\xxv$ of length $n$,
$\xzv \gets \fpsum{\xxv}$
computes a vector $\xzv$ of length $n$
such that $\xzvk{i} = \sum_{j=1}^i \xxvk{j}$ for $i \in [1,n]$.
We also define an inverse operation $\fpsuminv{}$.
Let $\xzv \gets \fpsuminv{\xxv}$ denote an operation
that computes $\xzv$ such that $\xxv = \fpsum{ \xzv }$.
This can be easily computed as
$\xzvk{1} := \xxvk{1}$ and $\xzvk{i} := \xxvk{i} - \xxvk{i-1}$ for all $i\in[2,n]$.
We further define reverse-ordered versions of these operations.
Given a vector $\xxv$ of length $n$,
$\xzv \gets \fpsumr{\xxv}$
computes a vector $\xzv$ of length $n$
such that $\xzvk{i} = \sum_{j=i}^n \xxvk{j}$ for $i \in [1,n]$.
Given a vector $\xxv$ of length $n$,
$\xzv \gets \fpsumrinv{\xxv}$
computes a vector $\xzv$ of length $n$
such that $\xxv = \fpsumr{\xzv}$.
This is computed as
$\xzv[n] := \xxv[n]$ and $\xzv[i] := \xxv[i] - \xxv[i+1]$ for all $i\in[1,n)$.

The protocol for group-wise sum is shown in \cref{alg:gb-sum}.
Let $r$ be the number of groups in the input.
In \crefrange{step:gb-sum-psumr}{step:gb-sum-psumri},
we compute $\xsviE{1}$ so that for each $j\in[1,r]$,
$\xsvi{1}[j]$ is the sum in the $j$-th group in $\xxv$.
This follows from the fact that
the collection of the first elements of each group in $\xpv$ is equal to
the reverse-ordered prefix sum of the sums in each group in $\xxv$.
Next, in \crefrange{step:gb-sum-copy-psuminv}{step:gb-sum-copy-psum},
we copy each $\xsvi{1}[j]$, which is the sum in the $j$-th group of $\xxv$, to each element of the $j$-th group.
To do this, we apply the technique used by Laud in his parallel reading protocol~\cite{Lau15} as follows.
We use the fact that when the prefix sum is computed,
an element with a value of zero will be copied with the result of the preceding element.
Specifically, in \cref{step:gb-sum-copy-unapply}, the values are restored to their original order
so that the first element of each group becomes the sum of each group
and the other elements become zero, and in \cref{step:gb-sum-copy-psum}, the prefix sum of the entire vector is computed.
However, this will copy the prefix sum of the sums instead of the sums.
Therefore, the inverse operation for the prefix sum is preliminarily performed in \cref{step:gb-sum-copy-psuminv}.

Note that this protocol is also useful for copying a particular element in a group
to all elements in the group, i.e., we clear all but the source elements with zeros and then apply this protocol.
This technique will be used in the following two protocols \cref{alg:gb-psum,alg:gb-max}.

The protocol uses $O(\xn \log \xn)$ non-free operations in $O(\log \xn)$ rounds.

\subsection{Our protocol for group-wise prefix sum}

\begin{algorithm}[htb]
 \caption{Group-wise prefix sum.}
 \label{alg:gb-psum}
 \algorithmcommon
 \KwNotation{$\xyvE \gets \fgbpsum{\xgvE, \xxvE}$.}
 \KwIn{A private \GFV{} $\xgvE$ of length $\xn$ and
       a private \IGV{} $\xxvE$ of length $\xn$.}
 \KwOut{A private vector $\xyvE$ of length $\xn$.}
 \KwComp{$O(\xn \log \xn)$ non-free operations in $O(\log \xn)$ rounds.}

 $\xsvE \gets \fpsum{\xxvE}$.
   \label{step:gb-psum-1}
 \;
 $\xqvkE{1} \gets 0$ and $\xqvkE{i} \gets \xsvkE{i-1} \times \xgvkE{i}$ for $i\in[2,\xn]$.
   \label{step:gb-psum-2}
 \;
 $\xyvE \gets \xsvE - \fgbsum{ \xgvE, \xqvE }$.
   \label{step:gb-psum-3}
 \;
\end{algorithm}

The {\em group-wise prefix sum} protocol privately computes prefix sums of each group in our data structure.
It receives a private \GFV{} $\xgvE$ of length $n$
and a private \IGV{} $\xxvE$ of values to be summed,
and outputs a private vector $\xyvE$ of prefix sums for each group
such that $\xyvk{i} = \sum_{j=\fhead{\xgv,i}}^i \xxvk{j}$.
An example of input/output is shown in \cref{fig:group-wise-io}.
Columns 1 and 2 are the inputs, and column 4 is the output.

The protocol is shown in \cref{alg:gb-psum}.
We first compute $\xsvE$, which is the prefix sum of $\xxvE$ (\cref{step:gb-psum-1}).
This looks almost done, but each value $\xsvk{i}$ exceeds the desired value
by a partial sum from the first element of $\xxv$ to the last element of the preceding group in $\xxv$.
Therefore, we try to subtract these partial sums from $\xsvE$ and obtain the desired output.
The predecessor of the first element in the $j$-th group in $\xsv$
is equal to
the partial sum from the first element in $\xxv$ to the last element in the $(j-1)$-th group of $\xxv$.
Using this property, we construct a vector $\xqvE$
that contains such values as the first values of the groups (\cref{step:gb-psum-2}).
We then copy the first elements of each group in $\xqvE$ to other elements by applying $\fgbsum{}$ protocol to $\xqvE$.
Finally, we subtract this from $\xsvE$ to obtain the prefix sum for each group (\cref{step:gb-psum-3}).

The protocol uses $O(\xn \log \xn)$ non-free operations in $O(\log \xn)$ rounds.

\subsection{Our protocol for group-wise max}

\begin{algorithm}[htb]
 \caption{Group-wise max.}
 \label{alg:gb-max}
 \algorithmcommon
 \newcommand{\txa}{\xxvjkE{(d)}{j-w}}
 \newcommand{\txb}{\xxvjkE{(d)}{j}}
 \newcommand{\tga}{\xgvjkE{(d)}{j-w}}
 \newcommand{\tgb}{\xgvjkE{(d)}{j}}
 \KwNotation{$\xyvE \gets \fgbmax{\xgvE, \xxvE}$.}
 \KwIn{A private \GFV{} $\xgvE$ of length $\xn$ and
       a private \IGV{} $\xxvE$ of length $\xn$.}
 \KwOut{A private vector $\xyvE$ of length $\xn$.}
 \KwComp{$O(\xn \log \xn)$ non-free operations in $O(\log \xn)$ rounds.}

 $\xm := \ceil{\log \xn}$.
   \label{alg:gb-max-init}
   \;
 $\xgvjE{(0)} \gets \xgvE$ and $\xxvjE{(0)} \gets \xxvE$. \;
 \For{$\xd := 0$ \KwTo $\xm-1$}{
   \label{alg:gb-max-ite}
   $\xw := 2^\xd$. \;
   $\xgvjE{(\xd+1)} \gets \xgvjE{(\xd)}$, $\xxvjE{(\xd+1)} \gets \xxvjE{(\xd)}$. \;
   \ForEachDIP{$j \in [\xw+1, \xn]$}{
     $\xgvjkE{(\xd+1)}{j} \gets \tga \bOR \tgb$. \;
     $\xavE \gets \fmax{\txa,\txb}$. \;
     $\xxvjkE{(\xd+1)}{j} \gets \fifelse{\tgb, \txb, \xavE}$.\label{alg:gb-max-end-loop}\;
   }
 }
 $\xtvkE{i} \gets \xgvkE{i+1}$ for $i\in[1,\xn)$ and $\xtvkE{\xn} \gets 1$.
   \label{alg:gb-max-tail}
 \;
 $\xyvE \gets \fgbsum{\xgvE, \xtvE \times \xxvjE{(\xm)}}$.
   \label{alg:gb-max-copy} \;
\end{algorithm}

The {\em group-wise max} protocol privately computes maximum values of each group in our data structure.
It receives a private \GFV{} $\xgvE$ of length $n$
and a private \IGV{} $\xxvE$ of length $n$,
and then outputs a private vector $\xyvE$ of the maximum values for each group such that $y[i] = \max_{ j\in[\fhead{\xgv,i}, \ftail{\xgv,i}] } x[j]$.
An example is shown in \cref{fig:group-wise-io}.
Columns 1 and 2 are the inputs, and column 5 is the output.

The protocol is shown in \cref{alg:gb-max}.
First, in \crefrange{alg:gb-max-init}{alg:gb-max-end-loop},
we compute $\xxvjE{(m)}$, where $j$-th element in $\xxvj{(m)}$ represents the maximum value up to $\xxv[j]$ in the group in $\xxv$.
That is, $\xxvj{(m)}[j] = \max_{i \in [\fhead{\xgv,j}, j]} \xxv[i]$.
The underlying idea is as follows.
Suppose we have a vector $\xxvi{w}$ that satisfies $\xxvi{w}[i] = \max_{j\in(i-w,i]}\xxv[j]$ for each $i$.
Then, we can obtain $\xxvi{2w}$ by computing $\xxvi{2w}[i]:=\max(\xxvi{w}[i], \xxvi{w}[i-w])$ for each $i$.
Since $\xxvi{1} = \xxv$, we can compute $\max_{j\in[1,i]}\xxv[j]$ for each $i$ by repeating this for $\Theta(\log n)$ rounds.
In the group-by prefix sum protocol, we want to compute $\max_{j\in[\fhead{\xgv,i},i]}\xxv[j]$ instead of $\max_{j\in[1,i]}\xxv[j]$.
Therefore, in addition to the maximum value in the range $(i-w,i]$, we keep a flag indicating whether the first element of the group is in this range or not.
Then, we can compute the desired value by not updating the maximum value if the corresponding flag is $1$.
Then, in \cref{alg:gb-max-tail,alg:gb-max-copy},
we copy the last elements of each group in $\xxvjE{(m)}$ to other elements as we have done in \cref{alg:gb-psum}.
Here, $\xtv[i] = 1$ if and only if the $i$-th element is the last element in a group.

The protocol uses $O(\xn \log \xn)$ non-free operations in $O(\log \xn)$ rounds,
since each iteration in \cref{alg:gb-max-ite} requires $O(\xn)$ non-free operations in $O(1)$ rounds.

\section{Our efficient decision tree training protocol}
\label{sec:our-dt-training}

In this section, we present our decision tree training protocol.
Given a private training data set, it outputs the trained decision tree in a private form.
Since the output decision tree is normalized for efficiency, we first describe it in \cref{sec:dt-normalization}.
Then, in \cref{sec:dttrainprotocol}, we explain how our protocol trains the tree in a layer-by-layer manner.
This part contains the main idea to reduce the $2^h$ factor in the computational cost to $h$.
The details of the batch \TEST{} selection are deferred to \cref{sec:test-selection}.
Note that our group-wise operations described in \cref{sec:group-ops} are used throughout our protocols presented in this section.

\subsection{Decision tree normalization for efficiency}
\label{sec:dt-normalization}
Our training protocol outputs an equivalent normalized decision tree instead of the one that should have been obtained when training in the clear.
Equivalent in this case means that the output for any given input is the same.
Roughly speaking, our normalization aligns the heights of all leaf nodes to the upper bound on the tree height by inserting internal nodes that forward any sample to the false child node.
Although this increases the number of nodes in the tree, in MPC, it reduces the data size and computational cost (though by just a constant factor),
and simplifies the protocol.

Before describing the details of our normalization,
let us recall the decision tree we originally wanted to compute, which is the output of \cref{alg:dt-train-plain}.
It is a binary tree with height less than or equal to $\xh$, and all \TEST{}s for its internal nodes are of the form \testa{j}{t},
where $j$ is an attribute number and $t$ is a threshold.

Our normalization consists of two modifications.
The first modification is to change each \TEST{} \testa{j}{t} to an equivalent \TEST{} \testb{j}{t'} such that $t' = 2t$.
In the original algorithm, we compute $t = (\xxvi{j}[i] + \xxvi{j}[i+1]) / 2$ when computing a threshold,
but this involves division, which is a costly operation in MPC.
We avoid division by computing $t' = 2t = \xxvi{j}[i] + \xxvi{j}[i+1]$ instead of $t$.

Our second modification is to align the height of all leaf nodes without changing the tree's output.
The modification is simple.
We insert an internal node $u$, which does not actually split, into the position of a leaf node $v$ with height less than $\xh$.
The \TEST{} for $u$ is \testb{1}{\cMINVAL}, which always returns false, and $u$ has a false branch to $v$,
where $\cMINVAL$ is a sufficiently small public value.
Any input tuple that reaches $u$ passes through the false branch to reach $v$, so the predicted label does not change.
In the normalized tree, all nodes with height less than $\xh$ are internal nodes, and all nodes with height $\xh$ are leaf nodes.
This makes our protocol simple and efficient.

\subsection{Our layer-by-layer training protocol}
\label{sec:dttrainprotocol}

\begin{algorithm*}[htb]
 \caption{Decision tree training.}
 \label{alg:train}
 \algorithmcommon
 \KwNotation{$\seq{ \xLjE{(\xk)} }_{\xk\in[0,\xh]} \gets \ftrainU{ \seq{ \xxvijE{j}{(0)} }_{j\in[1,\xm]}, \xyvjE{(0)}{}, \xh }$}
 \KwIn{
   $\xm$ private vectors
   $\seq{ \xxvijE{j}{(0)} }_{j\in[1,\xm]}$
   of length $\xn$,
   a private vector $\xyvjE{(0)}{}$ of length $\xn$, and
   an integer $\xh$.}
 \KwOut{
   A private decision tree 
   $\seq{ \xLjE{(\xk)} }_{\xk\in[0,\xh]}$
   of height $\xh$.
 }
 \KwComp{$O(\xh\xm\xn \log \xn)$ non-free operations in $O(\xh(\log \xn + \log \xm))$ rounds.}

 $\xgvjkE{(0)}{1} \gets 1$ and $\xgvjkE{(0)}{i} \gets 0$ for $i\in[2,\xn]$.
   \label{step:dt-train-init-g}
 \;
 $\xNjkE{(0)}{i} \gets 1$ for $i\in[1,\xn]$.
   \label{step:dt-train-init-N}
 \;
 \For{$\xk := 0$ \KwTo $\xh-1$}{
   \label{step:dt-train-loop}
   $\xLjE{(\xk)}, \xbvjE{(\xk)} \gets \fInNode{\xk, \seq{ \xxvijE{j}{(\xk)} }_j, \xyvjE{(\xk)}, \xgvjE{(\xk)}, \xNjE{(\xk)} }$.
     \label{step:dt-train-innode}
   \;
   $\xNE \gets  2^{\xk} \times \xbvjE{(\xk)} + \xNjE{(\xk)}$.
     \label{step:dt-train-next-N}
   \;
   $\xgvE \gets \fFirstOne{ \xgvjE{(\xk)}, \lnot\xbvjE{(\xk)} } + \fFirstOne{ \xgvjE{(\xk)}, \xbvjE{(\xk)} }$.
     \label{step:dt-train-next-g}
   \;
   $\xxvijE{1}{(\xk+1)}, \dots, \xxvijE{\xm}{(\xk+1)}, \xyvjE{(\xk+1)}, \xgvjE{(\xk+1)}, \xNjE{(\xk+1)} \gets \fsort{ \xbvjE{(\xk)}; \xxvijE{1}{(\xk)}, \dots, \xxvijE{\xm}{(\xk)}, \xyvjE{(\xk)}, \xgvE, \xNE }$.
     \label{step:dt-train-sort}
   \;
 }
 $\xLjE{(\xh)} \gets \fLeafNode{ \xh, \xgjE{(\xh)}, \xyvjE{(\xh)}, \xNjE{(\xh)} }$.
   \label{step:dt-train-layer-h}
 \;
\end{algorithm*}

\begin{algorithm}[htb]
 \caption{Detecting first ones in a private \IGV{}.}
 \label{alg:group-first-one}
 \algorithmcommon
 \KwNot{$\xfvE \gets \fFirstOne{\xgvE, \xbvE}$.}
 \KwIn{
   A private \GFV{} $\xgvE$ of length $\xn$ and
   a private bit vector $\xbvE$ of length $\xn$.}
 \KwOut{
   A private bit vector $\xfvE$ of length $\xn$.
 }
 \KwComp{$O(\xn \log \xn)$ non-free operations in $O(\log \xn)$ rounds.}

 $\xsvE \gets \fgbpsum{\xgvE, \xbvE}$.
 \;
 $\xfvE \gets \feq{ \xsvE \times \xbvE }{ 1 }$.
 \;
\end{algorithm}

This section describes the main part of our decision tree training protocol.
We construct a decision tree by training nodes of the same height in a batch, layer by layer, while keeping the input and output secret.
Training samples assigned to different nodes in the same layer
are processed as internally separate groups using the protocols proposed in \cref{sec:group-ops}.
This improves the $2^\xh$ factor of the communication complexity in \cite{AEV21} to $\xh$.

\subsubsection{Encoding of inputs and outputs}
Our decision tree training protocol receives a private training dataset
and a public upper bound $\xh$ on the height of the tree,
and outputs a private decision tree of height $\xh$.
The training dataset consists of $\xn$ samples, each of which consists of input tuple and a binary value called a class label.
Each input tuple consists of $\xm$ numerical input attribute values.
Our protocol receives it as
$\xm$ private vectors $\xxviE{j}$ ($j\in[1,\xm]$) of length $\xn$ and a private vector $\xyvE$ of length $\xn$.
That is, the $i$-th input tuple of the training dataset and
its associated class label correspond to $( \xxvik{1}{i}, \xxvik{2}{i}, \dots, \xxvik{\xm}{i} )$ and $\xyvk{i}$, respectively.

The output tree is a normalized binary tree as described in \cref{sec:dt-normalization}.
It is stored in $3\xh+2$ private vectors.
Since all nodes of height $\xk \in [0,\xh)$ are internal nodes, the node number,
attribute number, and threshold of each node are stored in three vectors
$\xNj{(\xk)}$, $\xAj{(\xk)}$, and $\xTj{(\xk)}$, respectively.
Since all nodes of height $\xh$ are leaf nodes,
the node number and \LL{} of each node are stored in two vectors
$\xNj{(\xk)}$ and $\xlj{(\xk)}$, respectively.
The length of each vector of height $\xk \in [0,\xh]$ is $\min\set{ \xn, 2^{\xk} }$.
If the actual number of nodes is smaller than the length of the vector, it is filled with a dummy value $\cNULL$.
The vectors of each layer are collectively denoted as
$\xLjE{(\xk)} := ( \xNjE{(\xk)}, \xAjE{(\xk)}, \xTjE{(\xk)} )$ ($\xk \in [0,\xh)$) and 
$\xLjE{(\xh)} := ( \xNjE{(\xh)}, \xljE{(\xh)} )$,
which we call {\em layer information}.

In order to represent the edges between nodes, each node is assigned an integer node number.
The only node with height $0$ is the root, and its node number is $1$.
For each child node (of height $\xk+1$) of a node of height $\xk$ with node number $d$,
assign node number $d$ to the false child (if any)
and node number $d+2^{\xk}$ to the true child (if any).
With this numbering scheme, in the $k$-th layer, all node numbers are assigned different values from $[1,2^{\xk}]$.

\subsubsection{The main protocol of our decision tree training}

The main protocol of our decision tree training is shown in \cref{alg:train}.
It trains the decision tree layer by layer in order from the $0$-th layer to the $\xh$-th layer.
Samples and associated values are stored in our \PDS{} as \IGV{}s.
Throughout the training, a \GFV{} $\xgvj{(\xk)}$ represents the grouping to each node of the $\xk$-th layer.
$\xNj{(\xk)}$,$\xxvij{j}{(\xk)}$, and $\xyvj{(\xk)}$ are \IGV{}s grouped by $\xgvj{(\xk)}$ that store
the node numbers, input attribute values, and output attribute values, respectively.
In the $0$-th layer, all samples are initialized to be assigned to the root node whose node number is $1$ (\cref{step:dt-train-init-g,step:dt-train-init-N}).
Then, each layer is trained iteratively (\cref{step:dt-train-loop}).

At each iteration,
we first trains the nodes at the $\xk$-th layer
and computes \BR{} $\xbvj{(\xk)}$ for each sample (\cref{step:dt-train-innode}).
This is executed in the $\fInNode{}$ protocol, which we will describe in \cref{sec:train-innode}.
Each $\xbvjk{(\xk)}{i}$ represents the \BR{} of the $i$-th sample.
0 and 1 denote false and true, respectively.

The node numbers $\xN$ and group flags $\xgv$ at the next layer are computed in \cref{step:dt-train-next-N,step:dt-train-next-g}, respectively.
Then, $\xgvj{(\xk+1)}$, $\xNj{(\xk+1)}$, $\xxvij{j}{(\xk+1)}$, and $\xyvj{(\xk+1)}$
of the $(\xk+1)$-th layer are computed by stably sorting
$\xgv$, $\xN$, $\xxvij{j}{(\xk)}$, and $\xyvj{(\xk)}$
 by $\xbvj{(\xk)}$ (\cref{step:dt-train-sort}).
Thanks to the stability of sorting, both the correspondence between the values of each sample and the contiguity of elements in the same group are maintained.

Let us verify the correctness of node numbers $\xN$ and group flags $\xgv$ for the next layer.
Let $v$ be a node at the $\xk$-th layer with node number $d$.
The node number of a child node of $v$ is $d$ for a false child and $d + 2^{\xk}$ for a true child.
Thus, $\xNE$, which is computed in \cref{step:dt-train-next-N}, is the node number in the next layer of each sample.
As for the group flags,
since the splitting of the groups is stable, the first 0 and 1 in each group are the first elements of the group after the split.
Since $\lnot\xbvjE{(\xk-1)}$ and $\xbvjE{(\xk-1)}$ indicate the positions of $0$'s and $1$'s, respectively,
the first elements of the new groups can be detected using $\fFirstOne{}$,
which detects first $1$'s for all groups.
We can construct $\fFirstOne{}$ by detecting elements whose value is $1$
and whose prefix sum in the group is also $1$, as shown in \cref{alg:group-first-one}.
The $\fFirstOne{}$ protocol uses $O(\xn\log\xn)$ non-free operations in $O(\log\xn)$ rounds.

In \cref{step:dt-train-layer-h},
we train the leaf nodes in a batch by invoking the $\fLeafNode{}$ protocol, which we will describe in \cref{sec:train-leaf}, and obtain the output vectors $\xLj{(\xh)}$ for height $\xh$.

The decision tree training protocol uses $O(\xh\xm\xn\log\xn)$ non-free operations in $O(\xh(\log\xm+\log\xn))$ rounds,
since
the $\fInNode{}$ protocol uses $O(\xm\xn \log \xn)$ non-free operations in $O(\log \xn + \log \xm)$, and
the $\fLeafNode{}$ protocol uses $O(\xn \log \xn)$ non-free operations in $O(\log \xn)$ rounds,
as we will show in the following sections.

\subsubsection{Batch training for internal nodes}
\label{sec:train-innode}

\begin{algorithm*}[htb]
 \caption{Training internal nodes.}
 \label{alg:train-internal-node}
 \algorithmcommon
 \KwNotation{$\xLjE{(\xk)}, \xbvE \gets \fInNode{\xk, \seq{ \xxviE{j} }_{j\in[1,\xm]}, \xyvE, \xgvE, \xNE }$.}
 \KwIn{
   An integer $\xk$,
   $\xm$ private vectors $\seq{ \xxviE{j} }_{j\in[1,\xm]}$ of length $\xn$,
   a private vector $\xyvE$ of length $\xn$,
   a private \GFV{} $\xgvE$ of length $\xn$, and
   a private vector $\xNE$ of length $\xn$.}
 \KwOut{
   $\xLjE{(\xk)} = (\xNjE{(\xk)}, \xAjE{(\xk)}, \xTjE{(\xk)})$ and
   a private bit vector $\xbvE$ of length $\xn$,
   where $\xNjE{(\xk)}$, $\xAjE{(\xk)}$, and $\xTjE{(\xk)}$ are
   private vectors of length $\min\set{\xn, 2^{\xk}}$.
 }
 \KwComp{$O(\xm\xn \log \xn)$ non-free operations in $O(\log \xn + \log \xm)$ rounds.}
 $\xAE, \xTE \gets \fsplitU{ \seq{ \xxviE{j} }_j, \xyvE, \xgvE }$.
   \label{step:in-gts}
 \;
 $\xsvE \gets \fGroupSame{\xgvE, \xyvE}$.
   \label{step:in-same}
 \;
 $\xAE, \xTE \gets \fifelse{ \xsvE ; 1, \cMINVAL; \xAE, \xTE }$.
   \label{step:in-overwrite}
 \;
 $\xbvE \gets \feval{ \seq{ \xxviE{j} }_j, \xAE, \xTE }$.
   \label{step:in-applytests}
 \;
 $\xNjE{(\xk)}, \xAjE{(\xk)}, \xTjE{(\xk)} \gets \fLayerShrink{ \xk, \xgvE, \xNE, \xAE, \xTE }$.
   \label{step:in-shrink}
 \;
\end{algorithm*}

\begin{algorithm}[htb]
 \caption{Checking if all elements in each group are the same.}
 \label{alg:samey}
 \algorithmcommon
 \KwNot{$\xfvE \gets \fGroupSame{\xgvE, \xyvE }$.}
 \KwIn{
   A private \GFV{} $\xgvE$ of length $\xn$ and
   A private bit vector $\xyvE$ of length $\xn$.
 }
 \KwOut{
   A private bit vector $\xfvE$ of length $\xn$.
 }
 \KwComp{$O(\xn \log \xn)$ non-free operations in $O(\log \xn)$ rounds.}
 $\xsvE \gets \fgbsum{ \xgvE, \vec{1} }$, where $\vec{1}$ is a vector $\seq{1,\dots,1}$ of length $\xn$. \;
 $\xsviE{0} \gets \fgbsum{ \xgvE, \lnot\xyvE }$. \;
 $\xsviE{1} \gets \fgbsum{ \xgvE, \xyvE }$. \;
 $\xfvE \gets \feq{\xsvE}{\xsviE{0}} \bOR \feq{\xsvE}{\xsviE{1}}$. \;
\end{algorithm}

\begin{algorithm*}[htb]
 \caption{Applying tests to samples.}
 \label{alg:eval}
 \algorithmcommon
 \KwNot{$\xbvE \gets \feval{ \seq{ \xxviE{j} }_{j\in[1,\xm]}, \xAE, \xTE }$}
 \KwIn{
   $\xm$ private vectors $\seq{ \xxviE{j} }_{j\in[1,\xm]}$ of length $\xn$,
   a private vector $\xAE$ of length $\xn$, and
   a private vector $\xTE$ of length $\xn$.
 }
 \KwOut{
   A private bit vector $\xbvE$ of length $\xn$.
 }
 \KwComp{$O(\xm\xn)$ non-free operations in $O(1)$ rounds.}

 $\xeviE{j} \gets \feq{ \xAE }{j }$ \forj.
   \label{step:eval-1}
 \;
 $\xxvkE{i} \gets \sum_{j\in[1,\xm]} \xxvikE{j}{i}\times\xevikE{j}{i}$ \fori.
   \label{step:eval-2}
 \;
 $\xbvE \gets \flt{ 2 \times \xxvE }{ \xTE }$.
   \label{step:eval-3}
 \;
\end{algorithm*}

\begin{algorithm*}[htb]
 \caption{Formatting vectors for $\xk$-th layer.}
 \label{alg:layershrink}
 \algorithmcommon
 \KwNot{
   $\seq{ \xdviE{1}, \dots, \xdviE{\xc} } \gets \fLayerShrink{ \xk, \xgvE, \xaviE{1}, \dots, \xaviE{\xc} }$.}
 \KwIn{
   An integer $\xk$,
   a private \GFV{} $\xgvE$ of length $\xn$,
   and $\xc$ private vectors $\xaviE{1}, \dots, \xaviE{\xc}$ of length $\xn$.
 }
 \KwOut{
   A sequence of
   $\xc$ private vectors $(\xdviE{1}, \dots, \xdviE{\xc})$, where length of $\xdviE{j}$ is  $\min\set{2^{\xk},\xn}$ for $j\in[1,c]$.
 }
 \KwComp{$O(c \xn \log \xn)$ non-free operations in $O(\log \xn)$ rounds.}

 $\xvviE{j} \gets \fifelse{ \xgvE, \xaviE{j}, \cNULL }$ for $j\in[1,\xc]$.
 \;
 $\xvviE{1}, \dots, \xvviE{\xc} \gets \fsort{ \lnot\xgvE; \xvviE{1}, \dots, \xvviE{\xc} }$.
 \;
 Let $\xdviE{j}$ be the first $\min\set{2^{\xk},\xn}$ elements of $\xvviE{j}$ for $j\in[1,\xc]$.
 \;
\end{algorithm*}

This section describes the protocol for training internal nodes in a batch, which we have been putting off.
It receives the privately grouped dataset of the $\xk$-th layer ($\xk\in[0,\xh)$),
computes the best test for each node,
and outputs the test results $\xbvE$ and the layer information $\xLjE{(\xk)} = (\xNjE{(\xk)}, \xAjE{(\xk)}, \xTjE{(\xk)})$.

In \cref{step:in-gts}, we compute the best test for each group using the $\fsplitU{}$ protocol which will be shown in \cref{sec:GTS}.
In \cref{step:in-same}, we determine if $\xyv[i]$ is all the same in each group, which is part of the stopping criterion described in \cref{sec:typical-dt-train}.
It is computed by the $\fGroupSame{}$ protocol shown in \cref{alg:samey}.
In the $\fGroupSame{}$ protocol, $\xsv$, $\xsvi{0}$, and $\xsvi{1}$ represent
the number of elements, the number of $0$, and the number of $1$ in the group, respectively.
Thus, $\feq{\xsv}{\xsvi{0}} \bOR \feq{\xsv}{\xsvi{1}}$ computes the desired value.
It uses $O(\xn \log \xn)$ non-free operations in $O(\log \xn)$ rounds.
In \cref{step:in-overwrite}, we replace \TEST{} with \testb{1}{\cMINVAL} for each element whose result in \cref{step:in-same} is true.
Specifically, for each $i$ such that $s[i]=1$, $\xAkE{i}$ and $\xTkE{i}$ are overwritten with $1$ and $\cMINVAL$, respectively.
In \cref{step:in-applytests}, the $\feval{}$ protocol is used to compute the test results from the best tests computed in the \cref{step:in-gts}.
In \cref{step:in-shrink}, the $\fLayerShrink{}$ protocol is used to format $\xNE$, $\xAE$, and $\xTE$.

The $\feval{}$ protocol is shown in \cref{alg:eval}.
It computes the private results $\xbvE$ of applying the tests denoted by $\xA$ and $\xT$ to the input tuples $\seq{ \xxvi{j} }_{j\in[1,\xm]}$.
For each $i \in [1,\xn]$, it computes the flag $\xeikE{j}{i}$ indicating whether $\xA[i] = j$ by an equality test (\cref{step:eval-1}),
and uses it to compute the $\xAk{i}$-th attribute value $\xxvi{\xAk{i}}[i]$ (\cref{step:eval-2}).
Then, it computes the \TEST{} result of each element in \cref{step:eval-3}.
It uses $O(\xm\xn)$ non-free operations in $O(1)$ rounds.

The $\fLayerShrink{}$ protocol is shown in \cref{alg:layershrink}.
It removes redundant values from given vectors.
Since the node numbers, attribute numbers, thresholds, and leaf labels are all the same
in the same group, it is sufficient to leave only one element for each group.
Therefore, we clear all but the first elements in each group with $\cNULL$ and move the first elements to the front of the vector.
Since the number of nodes in the $k$-th layer is at most $\min\set{n,2^k}$, we delete the trailing elements so that each vector has this length.
It uses $O(c\xn \log \xn)$ non-free operations in $O(\log \xn)$ rounds.

The $\fInNode{}$ protocol uses $O(\xm\xn \log \xn)$ non-free operations in $O(\log \xn + \log \xm)$ rounds,
since the $\fsplitU{}$ protocol uses $O(\xm \xn \log \xn)$ non-free operations in $O(\log \xn + \log \xm)$ rounds,
as we will show in \cref{sec:GTS}.

\subsubsection{Batch training for leaf nodes}
\label{sec:train-leaf}

\begin{algorithm*}[htb]
 \caption{Training leaf nodes.}
 \label{alg:train-leaf}
 \algorithmcommon
 \KwNotation{$\xLjE{(\xh)} \gets \fLeafNode{ \xh, \xgvE, \xyvE, \xNE }$.}
 \KwIn{
   An integer $\xh$,
   a private \GFV{} $\xgvE$ of length $\xn$,
   a private vector $\xyvE$ of length $\xn$,
   and a private vector $\xNE$ of length $\xn$.
 }
 \KwOut{
   $\xLjE{(\xh)} = (\xNjE{(\xh)}, \xljE{(\xh)})$,
   where $\xNjE{(\xh)}$ and $\xljE{(\xh)}$ are
   private vectors of length $\min\set{\xn, 2^{\xh}}$.
 }
 \KwComp{$O(\xn \log \xn)$ non-free operations in $O(\log \xn)$ rounds.}

 Computes the \LL{}s by
 $\xlE \gets \flt{ \fgbsum{ \xgvE, \lnot\xyvE } }{ \fgbsum{ \xgvE, \xyvE } }$,
 which are the most frequent values of $\xyv$ in each group. \;
    \label{step:train-leaf-ll}
 $\xNjE{(\xh)}, \xljE{(\xh)} \gets \fLayerShrink{ \xh, \xgvE, \xNE, \xlE }$. \;
   \label{step:train-leaf-shrink}
\end{algorithm*}

This section describes the protocol for training leaf nodes in a batch, which
we have putting off.
It receives the privately grouped dataset of the $\xh$-th layer, computes
the \LL{} for each node, and outputs the layer information $\xLjE{(\xh)} = (\xNjE{(\xh)}, \xljE{(\xh)})$.
The protocol is shown in \cref{alg:train-leaf}.
In \cref{step:train-leaf-ll}, we compute the most frequent value of output attribute as a \LL{}.
This is a typical method to define \LL{}.
In \cref{step:train-leaf-shrink}, the $\fLayerShrink{}$ protocol is used to format $\xNE$ and $\xlE$.

The protocol uses $O(\xn \log \xn)$ non-free operations in $O(\log \xn)$ rounds.

\subsection{Our batch test selection protocol}
\label{sec:test-selection}
In this section, we explain how to perform a batch node-wise \TEST{} selection
on a dataset that is grouped by nodes and stored in our \PDS{}.
Like the standard \TEST{} selection method in the clear, our batch \TEST{} selection consists of
three levels of components:
one to select the best \TEST{} across all attributes called the \GTS{} protocol,
one to select the best \TEST{} for splitting by a specific attribute called the \LTS{} protocol,
and one to compute measures called the \MGI{} protocol.
We will introduce them in order.
Thanks to our group-wise operations, all of them are almost straightforward to construct.

\subsubsection{Global test selection}
\label{sec:GTS}

\begin{algorithm*}[htb]
 \caption{Batch \GTS{}.}
 \label{alg:splitunsrt}
 \algorithmcommon
 \renewcommand{\xS}{\sigma}
 \KwNot{$\xavE, \xtvE \gets \fsplitU{ \seq{ \xxviE{j} }_{j\in[1,\xm]}, \xyvE, \xgvE }$}
 \KwIn{
   $\xm$ private vectors $\seq{ \xxviE{j} }_{j\in[1,\xm]}$ of length $\xn$,
   a private vector $\xyvE$ of length $\xn$,
   a private \GFV{} $\xgvE$ of length $\xn$.
 }
 \KwOut{
   A private vector $\xavE$ of length $\xn$ and
   a private vector $\xtvE$ of length $\xn$.
 }
 \KwComp{$O(\xm \xn \log \xn)$ non-free operations in $O(\log \xn + \log \xm)$ rounds.}

 \ForEachDIP{$j \in [1,\xm]$}{
   \label{step:splitunsrt-loop}
   $\xuviE{j}, \xvviE{j} \gets \fsort{ \fpsum{ \xgvE }, \xxviE{j}; \xxviE{j}, \xyvE }$.
     \label{step:splitunsrt-sort}
   \;
   $\xtviE{j}, \xsviE{j} \gets \faws{ \xgvE, \xuviE{j}, \xvviE{j} }$.
     \label{step:splitunsrt-aws}
   \;
 }
 \ForEachDIP{$i\in[1,\xn]$}{
   $\xavkE{i}, \xtvkE{i}  \gets \fvectmax{ \seq{ \xsvikE{1}{i}, \dots, \xsvikE{\xm}{i} }; \seq{ 1, \dots, \xm }, \seq{ \xtvikE{1}{i},\dots,\xtvikE{\xm}{i} } }$.
   \label{step:splitunsrt-argmax}
 \;
 }
\end{algorithm*}

The \GTS{} protocol computes the best test through all attributes for each node in a batch.
The algorithm is straightforward: it calls the \LTS{} protocol to compute the best test for each attribute and then selects the best test among them.
Since the \LTS{} protocol assumes that the data is already sorted within a group for a given attribute, this protocol is responsible for sorting within the group before calling \LTS{}.

The protocol is shown in \cref{alg:splitunsrt}.
It receives the training data (input tuples and class labels) privately grouped by nodes, and outputs the information (attribute number and threshold) of the best test for each group.
For each input attribute, it sorts the input attribute values and class labels within the group and selects the best test for each group when splitting on that attribute (\crefrange{step:splitunsrt-loop}{step:splitunsrt-aws}).
Then select the best test among all attributes in \cref{step:splitunsrt-argmax}.
Since the output of the \LTS{} protocol is identical in each group, it is sufficient to do this for each element independently.

The protocol is almost identical to the algorithm in the clear and the protocol in \cite{AEV21}.
The difference is that we need to sort within each group in \cref{step:splitunsrt-sort}.
Recalling that $\xgv$ is a bit vector where only the first element of each group is $1$, we can see that $\fpsum{\xgv}$ computes different and ascending values for each group.
Thus, we can sort within each group by using $\fpsum{\xgv}$ and $\xxvi{j}$ as keys in lexicographic order as in \cref{step:splitunsrt-sort}.

The protocol uses $O(\xm\xn \log \xn)$ non-free operations in $O(\log\xm+\log\xn)$ rounds,
since $\faws{}$ protocol uses $O(\xn \log \xn)$ non-free operations in $O(\log \xn)$ rounds,
as we will show in the next section.

\subsubsection{Attribute-wise test selection}

\begin{algorithm*}[htb]
 \caption{Batch \LTS{}.}
 \label{alg:test}
 \algorithmcommon
 \renewcommand{\xe}{\theta}

 \KwNot{$\xtvE, \xsvE \gets \faws{ \xgvE, \xxvE, \xyvE }$}

 \KwIn{%
 A private \GFV{} $\xgvE$ of length $\xn$,
 a private vector $\xxvE$ of length $\xn$, and
 a private vector $\xyvE$ of length $\xn$.}

 \KwOut{
   Private vectors $\xtvE$ and $\xsvE$ of length $\xn$.
 }
 \KwComp{$O(\xn \log \xn)$ non-free operations in $O(\log \xn)$ rounds.}

  $\xsvE \gets \fgini{\xg,\xy}$. \;
  \label{step:test-gini}
  $\xtvkE{i} \gets \xxvkE{i} + \xxvkE{i+1}$ for all $i\in[1,n)$ and
         $\xtvkE{n} \gets \cMINVAL$. \;
  \label{step:test-th}

  $\xpvkE{i} \gets \xgvkE{i+1} \bOR \feq{ \xxvkE{i}}{\xxvkE{i+1} }$ for $i\in[1,\xn)$ and $\xpvkE{\xn} \gets 1$. \;
  \label{step:skip1}
  $\xsvE, \xtvE \gets \fifelse{\xpvE; \cMINVAL, \cMINVAL; \xsvE, \xtvE}$. \;
  \label{step:skip2}
  $\xsvE, \xtvE \gets \fgbmax{\xgvE, \xsvE; \xsvE, \xtvE}$.
    \label{step:test-gbmax} \;
\end{algorithm*}

The \LTS{} protocol computes the best \TEST{}s in each group for a given numerical input attribute.
It assumes that the input attribute values and class labels are sorted with respect to the input attribute values within each group.
It implements the technique by Abspoel et al.~\cite{AEV21} that
reduces the number of candidate thresholds from $\Theta(n^2)$ to $\Theta(n)$,
on our data structure using the group-wise operations proposed in \cref{sec:group-ops}.
We use the $\fgini{}$ protocol, which will be described in \cref{sec:modified-gini}, to compute the modified Gini index.

The protocol is shown in \cref{alg:test}.
It receives a private \GFV{} $\xgvE$, a private numeric input attribute vector $\xxvE$, and a private class label vector $\xyvE$.
Vectors $\xxv$ and $\xyv$ are sorted with respect to $\xxv$ in each group.
Outputs are thresholds $\xtvE$ and \SCORE{}s $\xsvE$ of the best \TEST{}s in each group.

We show that the protocol computes the best \TEST{}s in each group for a given numerical input attribute.
Since $\xxv$ and $\xyv$ are sorted within each group with respect to $\xxv$,
it is sufficient to consider only the split between two adjacent elements in each group \cite{AEV21}.
In \cref{step:test-gini,step:test-th}, the threshold $\xtvk{i}$ and the \SCORE{} $\xsvk{i}$
for split between the $i$-th and $(i+1)$-th elements are computed.
If the $i$-th element is the last element in a group (i.e., $i=\xn$ or $\xgvk{i+1}=1$)
or if it has the same attribute value as the next element (i.e., $\xxvk{i} = \xxvk{i+1}$),
we cannot split between the $i$-th and $(i+1)$-th elements.
In this case, we set $\xtvk{i} := \cMINVAL$ and $\xsvk{i} := \cMINVAL$ (\cref{step:skip1,step:skip2}).
In \cref{step:test-gbmax}, the \SCORE{} and threshold of a element whose \SCORE{} is the maximum in a group are copied to other elements in the group.

The protocol uses $O(\xn\log\xn)$ non-free operations in $O(\log\xn)$ rounds,
since $\fgini{}$ protocol uses $O(\xn \log \xn)$ non-free operations in $O(\log \xn)$ rounds,
as we will show in the next section.

\subsubsection{Modified Gini index}
\label{sec:modified-gini}

\begin{algorithm*}[htb]
 \caption{Modified Gini index.}
 \label{alg:gini}
 \algorithmcommon
 \KwNot{$\xsvE \gets \fgini{\xgvE, \xyvE}$.}
 \KwIn{%
 A private \GFV{} $\xgvE$ of length $\xn$ and
 a private vector $\xyvE$ of length $\xn$.}
 \KwOut{%
 A private vector $\xsvE$ of length $\xn$,
 where $\xsv[i]$ is the modified Gini index when
 the dataset is split between the $i$-th and $(i+1)$-th elements.}
 \KwComp{$O(\xn \log \xn)$ non-free operations in $O(\log \xn)$ rounds.}

 $\xyviE{0} \gets \lnot\xyvE$, $\xyviE{1} \gets \xyvE$. \;
 $\xuviE{b} \gets \fgbpsum{\xgvE, \xyviE{b}}$ for $b \in \set{0,1}$. \;
 $\xsviE{b} \gets \fgbsum{\xgvE, \xyviE{b}}$ for $b \in \set{0,1}$. \;
 $\xwviE{b} \gets \xsviE{b} - \xuviE{b}$ for $b\in\set{0,1}$. \;
 $\xuvE \gets \xuviE{0} + \xuviE{1}$ and
 $\xwvE \gets \xwviE{0} + \xwviE{1}$. \;
 $\xpvE \gets \xwvE \times (\xuviE{0}^2 + \xuviE{1}^2) + \xuvE \times (\xwviE{0}^2 + \xwviE{1}^2)$. \;
 $\xqvE \gets \xuvE \times \xwvE$. \;
 $\xsvE \gets \xpvE / \xqvE$. Here, for simplicity, we set $\xsv$ to be the result of element-wise division of $\xpv$ and $\xqv$;
 however, in practice, we set $\xsv$ to be the pair of $\xpv$ and $\xqv$,
 and a comparison of the elements in $\xsv$ is replaced by the division-free comparison as \cite{AEV21}. \;
   \label{alg:gini-div}
\end{algorithm*}

We present a protocol to compute the modified Gini index for privately grouped dataset.
Thanks to the group-wise operations proposed in \cref{sec:group-ops},
the formula by Abspoel et al.~\cite{AEV21} in \cref{eq:modified-gini} can be used almost directly.

The protocol is shown in \cref{alg:gini}.
The input is a private \GFV{} $\xgvE$ and a private class label vector $\xyvE$ which is sorted by an input attribute in each group.
The output is a private vector $\xsvE$, where each $\xsvk{i}$ represents
the modified Gini index for a split between the $i$-th and $(i+1)$-th elements.

Since each $\xyvi{b}$ is a bit vector with $\xyvik{b}{i} = 1$ iff $\xyvk{i} = b$, let the $i$-th element be $v$, then $\xuvik{b}{i}$ represents the number of $b$ up to $v$ in the group, and $\xwvik{b}{i}$ represents the number of $b$ after $v$ in the group.
That is, $\xuvik{0}{i}$, $\xuvik{1}{i}$, $\xuvk{i}$, $\xwvik{0}{i}$, $\xwvik{1}{i}$, and $\xwvk{i}$ are the number of $0$'s up to $v$, the number of $1$'s up to $v$, the number of elements up to $v$, the number of $0$'s after $v$, the number of $1$'s after $v$, and the number of elements after $v$, respectively, in the group. 
From \cref{eq:modified-gini}, $\xpk{i} / \xqk{i}$ is the modified Gini index for splitting between the $i$-th and $(i+1)$-th elements.

The protocol uses $O(\xn\log\xn)$ non-free operations in $O(\log\xn)$ rounds.

\section{Demonstration of our protocol's practicality}

In order to show the practicality of our decision tree training protocol,
we implemented it and measured the running time.

\subsection{Implementation methods}

\begin{table}
 \centering
 \caption{Specifications of the machine used for our benchmark.}
 \label{tbl:environment}
 \begin{tabular}{ll}
OS      & CentOS Linux release 7.3.1611 \\
\hline
CPU     & Intel Xeon Gold 6144k (3.50GHz 8 core/16 thread)$\times$2 \\
\hline
Memory  & 768 GB \\
 \end{tabular}
\end{table}

We implemented our protocols on a Shamir's secret-sharing based three-party computation over a field $\setZ_p$, where $p = 2^{61}-1$.
This 3PC scheme is secure against a single static corruption.
For the ABB implementation, we used the comparison protocols by Kikuchi et al.~\cite{KIM+18} and the multiplication protocol by Chida et al.~\cite{CHI+19}.
We also replaced some of the protocols built on ABB with more efficient protocols.
The inner product, apply, unapply, and sortperm protocols are based on the ones by Chida et al.~\cite{CHI+19}.
Our implementation includes several optimizations.

The protocols were implemented with the C++ language.
We measured on three servers with the same configuration connected by a ring configuration of Intel X710/X557-AT 10G.
The configuration of the servers is shown in \cref{tbl:environment}.

\subsection{Benchmarking results}

\begin{table}
 \centering
 \caption{Running time for training a decision tree of different heights $h$ on $n=10^4$ samples with $m=10$ input variables.}
 \label{tbl:runtime-h}
 \begin{tabular}{rrrr}
  $h$ & Time [s] \\
  \hline
  1 & 1.342 \\
  2 & 2.203 \\
  5 & 4.432 \\
  10 & 8.810 \\
  20 & 16.633 \\
  50 & 40.891 \\
 \end{tabular}
\end{table}

\begin{table}
 \centering
 \caption{Running time for training a decision tree of height $h=5$ on different numbers of samples $n$ with $m = 10$ input variables.}
 \label{tbl:runtime-n}
 \begin{tabular}{rrrr}
  $n$ & Time [s] \\
  \hline
  $10$   & 1.811 \\
  $10^2$ &  2.092 \\
  $10^3$ &  2.569 \\
  $10^4$ &  4.432 \\
  $10^5$ &  32.035 \\
 \end{tabular}
\end{table}

\begin{table}
 \centering
 \caption{Running time for training a decision tree of height $h=5$ on $n=10^4$ samples with different numbers of input variables $m$.}
 \label{tbl:runtime-m}
 \begin{tabular}{rrrr}
  $m$ & Time [s] \\
\hline
   1 & 2.469 \\
   2 & 2.641 \\
   5 & 3.069 \\
   10 & 4.432 \\
   20 & 7.790 \\
   50 & 19.111 \\
   100 & 39.401 \\
 \end{tabular}
\end{table}

To show the scalability of our protocol, we measured the running time for different parameters of $n$, $m$, and $h$.
Based on the case where $n = 10^4$, $m = 10$, and $h = 5$, we measured the execution time of training by varying $n = 10, 10^2, 10^3, 10^4, 10^5$, $m = 1, 2, 5, 10, 20, 50, 100$, and $h = 1, 2, 5, 10, 20, 50$.
The results are shown in \cref{tbl:runtime-n,tbl:runtime-m,tbl:runtime-h}.
Each runtime is the average value of three measurements.
The results show that the running time is approximately linear with respect to $n$, $m$, and $h$, respectively.

%\bibliographystyle{jalpha}
%\bibliography{odt}
\newcommand{\etalchar}[1]{$^{#1}$}

\end{document}